	\documentclass[floatfix,prl,twocolumn,superscriptaddress]{revtex4-1}

\usepackage{amssymb}
\usepackage{amsmath}
\usepackage[dvips]{graphicx}
\graphicspath{{./figures/}}
\usepackage{bm}
\usepackage[english]{babel}
\usepackage{float}
\usepackage{color}
\usepackage{mathtools}

\newcommand{\be}{\begin{equation}}
\newcommand{\ee}{\end{equation}}
\newcommand{\bea}{\begin{eqnarray}}
\newcommand{\eea}{\end{eqnarray}}

\begin{document}

\title{Quasiparticle mass enhancement as a measure of entanglement in the Kondo problem}

\author{Nayra A. \'Alvarez Pari}
\affiliation{Centro At{\'o}mico Bariloche and Instituto Balseiro, CNEA, 8400 Bariloche, Argentina}
\author{D. J.  Garc\'ia}
\affiliation{Centro At{\'o}mico Bariloche and Instituto Balseiro, CNEA, 8400 Bariloche, Argentina}
\affiliation{Consejo Nacional de Investigaciones Cient\'{\i}ficas y T\'ecnicas (CONICET), Argentina}
\author{Pablo S. Cornaglia}
\affiliation{Centro At{\'o}mico Bariloche and Instituto Balseiro, CNEA, 8400 Bariloche, Argentina}
\affiliation{Consejo Nacional de Investigaciones Cient\'{\i}ficas y T\'ecnicas (CONICET), Argentina}

\begin{abstract}
	We analyze the quantum entanglement between opposite spin projection electrons in the ground state of the Anderson impurity model. In this model, a single level impurity with intralevel repulsion $U$ is tunnel coupled 
	to a free electron gas. The Anderson model presents a strongly correlated many body ground state with mass enhanced quasiparticle excitations.
	We find, using both analytical and numerical tools, that the quantum entanglement between opposite spin projection electrons is a monotonic universal function of the quasiparticle mass enhancement $Z$ in the Kondo regime.
	This indicates that the interaction induced mass enhancement, which is generally used to quantify correlations in quantum many body systems, could be used as a measure of entanglement in the Kondo problem.
	


%
%
%
%

\end{abstract}


\maketitle
Entanglement is a characteristic trait of quantum mechanics and a fundamental resource for quantum information processing protocols. It is also a powerful tool to analyze interacting many-body systems, able to detect and characterize quantum phase transitions and topological phases \cite{osterloh2002scaling,osborne2002entanglement,gu2004entanglement,bayat2014order}, and plays a fundamental role in the thermalization process \cite{deutsch1991quantum,*srednicki1994chaos,*rigol2008thermalization}. 
Quantifying quantum entanglement in many body systems is, however, an experimentally difficult task \cite{islam2015measuring,*kaufman2016quantum,*lukin2019probing}. 

In this Letter we show that there is a one to one correspondence linking the quantum entanglement between opposite spin projection electrons and the interaction induced quasiparticle mass enhancement in the Kondo correlated many-body ground state of the Anderson impurity model.
The Anderson model describes a single level impurity with intralevel repulsion $U$ tunnel coupled to a free conduction electron band. 
It has been extensively analyzed, together with other quantum impurity problems, to describe diluted magnetic impurities in a metallic host \cite{hewson1997kondo,wilson1975kondo,georges2015beauty}, electronic transport through quantum dots \cite{costi2001magnetotransport,cornaglia2003transport}, and to solve models of strongly correlated electron materials using the self-consistent dynamical mean field theory (DMFT) equations \cite{georges1996dynamical}. Its most salient features are associated with the magnetic moment behavior at the impurity and the crossover to a spin-singlet low temperature behavior.
The Kondo effect, the screening of the local magnetic moment at the impurity by the conduction electrons, leads to a strongly correlated spin-singlet many-body ground state that dominates the physics below a characteristic Kondo temperature $T_K$. For $T<T_K$ the low-energy properties, as the impurity contribution the specific heat or the impurity magnetization at low magnetic fields, are universal functions of the relevant energy scale divided by $k_B T_K$.

Nozi\`eres successfully applied the Fermi liquid concept to analyze the low energy excitations of the Anderson model above the 
Kondo singlet ground state \cite{nozieres1974fermi}. 
Fermi liquid theory is based on the assumption of a one to one correspondence between the low energy excitations of an interacting electron system and those of a noninteracting Fermi gas (see, e.g. Refs. \cite{fulde2012correlated,hewson1997kondo}). It allows to describe the properties of a many-body electron system through an effective theory of weakly interacting quasiparticle excitations. The quasiparticles have a renormalized mass $m^\star =m/Z$, where $Z$ is the interaction induced quasiparticle mass enhancement, and $m$ the effective electron mass in the absence of electron-electron interactions in the conduction band. The quasiparticle mass enhancement $0<Z<1$ is generally used to quantify electron-electron correlations and the coherence scale in Fermi liquid systems \cite{masa1998mit}. In the Kondo problem, $Z\sim \pi k_B T_K/N\Gamma(\varepsilon_F)$, where the hybridization function $\Gamma(\varepsilon)$ characterizes the coupling between the impurity and the conduction electrons, $\varepsilon_F$ is the Fermi energy, and $N$ is the impurity level degeneracy \cite{hewson1997kondo}.

\begin{figure}[tb]\center
\includegraphics[width=\columnwidth,angle=0,keepaspectratio=true]{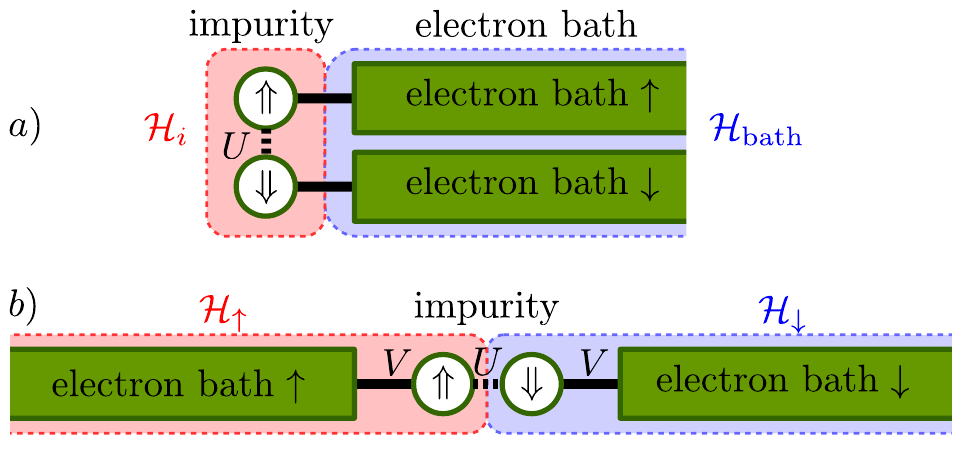}
	\caption{Sketch of the Anderson model, for an impurity with Krammers degeneracy, including the two partitions of the total Hilbert space $\mathcal{H}$ considered to calculate the entanglement entropy: a) $\mathcal{H}=\mathcal{H}_i\otimes\mathcal{H}_{\text{bath}}$. b) $\mathcal{H}=\mathcal{H}_\uparrow\otimes\mathcal{H}_\downarrow$. 
 The Coulomb repulsion $U$ between opposite spin electrons at the impurity and the tunnel coupling $V$ between the impurity level and the conduction electron bath, are indicated in the figure. 
	}
	\label{fig:partitions}
\end{figure}

We consider the ground state $|\Psi_{\text{GS}}\rangle$ of the Anderson model and use the von Neumann entropy to quantify the entanglement between two subspaces, $\mathcal{H}_A$ and $\mathcal{H}_B$, of the total Hilbert space $\mathcal{H}=\mathcal{H}_A\otimes \mathcal{H}_B$:
\begin{equation}
    S(\rho_A)=-\text{Tr}\left\{ \rho_A\log_2\rho_A \right\}=-\sum_i \lambda^A_i \log_2 \lambda^A_i,
    \label{eq:vn}
\end{equation}
where $\rho_A=\text{Tr}_B|\Psi_{\text{GS}}\rangle\langle\Psi_{\text{GS}}|$ is the partial trace over $\mathcal{H}_B$, and the $\lambda^A_i$ are the eigenvalues of $\rho_A$.
This entanglement measure has been successfully used to analyze the spatial extent of the correlations, i.e. the size of the Kondo screening cloud \cite{borzenets2020observation}, in a variety of Kondo models \cite{sorensen2007impurity,bayat2010negativity,bayat2012entanglement,alkurtass2016entanglement,Laflorencie20161,bayat2017scaling,feiguin2017} including systems showing quantum criticality \cite{PhysRevLett.121.147602}. In these works, the subspace $A$ corresponds to the impurity and a set lattice sites localized close to it, while subspace $B$ corresponds to the rest of the system [see Fig. \ref{fig:partitions}a)].
These partitions present quantum entanglement even for a noninteracting system ($U=0$). To analyze the entanglement generated by the Coulomb repulsion $U$ we focus our analysis on the partition between opposite spin projection electrons [see Fig. \ref{fig:partitions}b)] that are only coupled by the local interaction $U$ at the impurity.  The resulting entanglement entropy $S_\uparrow$ vanishes in the noninteracting limit.


In its simplest form, the Anderson model for an $N$ degenerate impurity reads $H=H_i+H_{cb}+H_V$ \cite{bickers1987review}:
where
\begin{align}
	H_i=&\sum_j \varepsilon_{j} f_{j}^\dagger f_{j}^{}+U\sum_{j^\prime>j}f_{j}^\dagger f_{j}^{}f_{j^\prime}^\dagger f_{j^\prime}^{},
\end{align}\label{eq:AM}
is the impurity Hamiltonian,
\begin{equation}
	H_{cb}=\sum_{kj}\varepsilon_{kj}c_{kj}^\dagger c_{kj}^{},
\end{equation}
models the conduction electron band, and
\begin{align}
	H_V=\sum_{kj}V_{k}\left(c_{kj}^\dagger f_{j}^{}+f_{j}^\dagger c_{kj}^{} \right),
\end{align}
models the tunnel coupling between the impurity and the conduction band.
Here $f_{j}^\dagger$ and $c_{kj}^\dagger$ create an electron, with quantum number $j$, at the impurity and at the conduction band level $k$, respectively. 
For $U=0$, assuming a constant density of states and neglecting the $k$ dependence of $V_k$ ($V_k=V$), the impurity level acquires a lifetime $\hbar /\Gamma$ and an associated spectral width $\Gamma=\pi \rho |V|^2$. 
For $U>0$ and $-U<\varepsilon_j<0$ the isolated impurity has a single occupancy and it can be regarded as magnetic impurity with a total angular momentum $J$. In the absence of an external magnetic field, we set $\varepsilon_{kj}=\varepsilon_{k}$ and $\varepsilon_{j}=\varepsilon$, and the degeneracy of the isolated impurity ground state is $N=2J+1$.   
In this parameter regime, the tunnel coupling to the conduction bath leads to the screening of the magnetic moment and to a singlet ground state.\\

\paragraph{\textit{The $U \to \infty$ and $N \to \infty$ Anderson model.\textemdash}} In the infinite-$U$ limit, the impurity multioccupancy is blocked (the impurity can be either empty or singly occupied) and to lowest order in $1/N$, the ground state is a singlet of the form \cite{varma1975,bickers1987review}
\begin{equation}\label{eq:GS}
	|\Psi_{\text{GS}}\rangle=a_0\left( |F\rangle +\frac{1}{\sqrt{N}}\sum_{kj} b_k f_{j}^\dagger  c_{kj}|F\rangle\right)
\end{equation}
where $|F\rangle=\prod_{\varepsilon_k\leq\varepsilon_F}\prod_{j=-J}^{J}c_{kj}^\dagger|0\rangle$ is the Fermi sea filled up to $\varepsilon_F$  and has associated an energy $E_0=N\sum_{\varepsilon_k\leq\varepsilon_F}\varepsilon_k$.
A variational calculation of the Kondo singlet energy $\varepsilon_K=E_0+\varepsilon-\langle\Psi_{\text{GS}}|H|\Psi_{\text{GS}}\rangle/\langle\Psi_{\text{GS}}|\Psi_{\text{GS}}\rangle$ leads to the equations (setting $\varepsilon_F=0$):
\begin{eqnarray}
	b_k=\frac{\sqrt{N} V_k}{-\varepsilon_K+\varepsilon_k}\\
	\varepsilon-\varepsilon_K= \frac{N}{\pi}\int_{-D}^{0}d\omega \frac{\Gamma(\omega)}{-\varepsilon_K +\omega},\label{eq:self}
\end{eqnarray}
where $\Gamma(\omega)=\pi\sum_k\delta(\omega-\varepsilon_k)V_k^2$, and $-D$ is the lowest energy of the conduction band. Deep in the Kondo regime ($-\varepsilon\gg N \Gamma$, $\varepsilon_K \ll D, |\varepsilon|$) the integral in Eq. (\ref{eq:self}) is dominated by the energies close to the Fermi level ($\omega\sim 0$) which allows to approximate $\Gamma(\omega)\sim\Gamma(0)=\Gamma$. This results in $\varepsilon_K\sim D e^{-\pi \varepsilon/N\Gamma}$. In the $N\to \infty$ limit, $N \Gamma$ is taken as constant, and Eq. (\ref{eq:GS}) is the exact ground state wavefunction \cite{bickers1987review}.

We calculate the entanglement entropy $S_\uparrow$ in the ground state wavefunction ($|\Psi_{\text{GS}}\rangle$) for the partition $\mathcal{H}_{j>0}\otimes \mathcal{H}_{j\leq 0}$ of the total Hilbert space. 
After an appropriate basis change (see Ref. \cite{suppmat}), the density matrix of the positive projection electrons, associated with nonzero eigenvalues can be written as 
\begin{equation}
	\rho_{j>0}=\text{Tr}_{j\leq 0}|\Psi_{\text{GS}}\rangle\langle\Psi_{\text{GS}}|=a_0^2\begin{pmatrix}1+\frac{b_0^2}{2}&\frac{b_0}{\sqrt{2}}\\\frac{b_0}{\sqrt{2}}&\frac{b_0^2}{2}\end{pmatrix},
\end{equation}
where $b_0^2=\frac{N\Gamma}{\pi} \int_{-D}^0 \frac{d\omega}{(-\varepsilon_K+\omega)^2}\sim \frac{N\Gamma}{\pi \varepsilon_K}$, and the normalization of the wavefunction leads to $a_0^2=(1+b_0^2)^{-1}$. The entanglement entropy $S_\uparrow$ can be readily calculated from the eigenvalues of $\rho_{j>0}$ using Eq. (\ref{eq:vn}). $S_\uparrow$ depends on the model parameters only through $Z=\pi \varepsilon_K/N\Gamma$ and it is a monotonic function of $Z$ (see Ref. \cite{suppmat}).  As a consequence, systems with different model parameters but the same $Z$ have the same spin entanglement entropy. Deep in the Kondo regime, $Z\ll1$ we have: 
\begin{equation}\label{eq:SZ0N}
  S_\uparrow\sim 1-\frac{Z}{\ln(2)}.
\end{equation}

The impurity-bath entanglement entropy $S_i$ does not lead to useful information on the nature of the correlations induced by $U$ in the large-$N$ limit. The impurity density matrix $\rho_i$ has $N$ eigenvalues equal to $n_f/N$, associated with the occupancies of the $N$ possible spin projections at the impurity and a single eigenvalue $1-n_f$ that corresponds to the empty state. Here $n_f=\langle\sum_j f^{\dagger}_{j}f^{}_{j}\rangle\leq 1$ is the ground state expectation value of the level occupancy. This leads to a diverging $S_i\sim n_f\log_2N$ in the large-$N$ limit, a behavior that is also obtained in the noninteracting case.\\

\paragraph{\textit{The finite $U$ and $N=2$ Anderson model.\textemdash}} To analyze the validity of the relation between the quasiparticle mass enhancement and the spin entanglement entropy $S_\uparrow$ in a more general case with finite $U$ and finite $N$, we resort to numerical calculations using the density matrix renormalization group (DMRG) \cite{peschel1999density,*hallberg2006new}. DMRG is a numerical method, based on Wilson's renormalization group ideas, to solve strongly correlated models in finite size systems. $m$ states are selected at each renormalization step according to their respective weight in the ground state wavefunction. The results are exact for large $m$, but $m$ is limited by the increase of computational cost. For the Anderson model, which can be mapped into a linear tight binding chain with the impurity at one end, the accuracy improves exponentially with $m$, and excellent results are obtained for $m<1000$ for a wide range of model parameters \cite{suppmat}.

We focus the numerical calculations on the $N=2$ ($J=1/2$) case and use the standard notation for the magnetic quantum number $j=\uparrow,\downarrow$ and the fermion operators $c_{f,j}\equiv f_{j}$.
To model the electron band, we consider a half-filled tight-binding chain of length $L$,
\begin{equation}
	H_{cb} = - t \sum_{i=1}^L\sum_{j=\{\uparrow,\downarrow\}} \left(c_{i,j}^\dagger c_{i+1,j}^{} + \text{H.c.}\right),
	\label{eq:tbchain}
\end{equation}	
which leads to a semielliptic local density of states $\rho(\varepsilon)=\frac{2}{\pi D^2}\sqrt{D^2-\varepsilon^2}$ at site $1$ for $L\to \infty$ and $t=D/2$. The tunnel coupling is given by
\begin{equation}
	H_{V} = \sum_{j=\{\uparrow,\downarrow\}}\left(V c_{f,j}^{\dagger} c_{1,j}^{} + \text{H.c.}\right). 
	\label{eq:hybchain}
\end{equation}
 The hybridization at the Fermi level ($\varepsilon_F=0$) is $\Gamma=\pi \rho(0)V^2=2 V^2/D$.

The reduced density matrices required to calculate the entanglement entropy can be obtained for finite $L$ using the DMRG \cite{peschel1999density,schollwock2005density}. We performed a finite size analysis for $L$ up to $4096$ which restricts the model parameters to regimes where $\varepsilon_K\gg t/L$ in order to avoid finite size effects \cite{thimm1999kondo,*cornaglia2002mesokondo,cornaglia2003transport}.

\begin{figure}[tb]\center
\includegraphics[width=\columnwidth,angle=0,keepaspectratio=true]{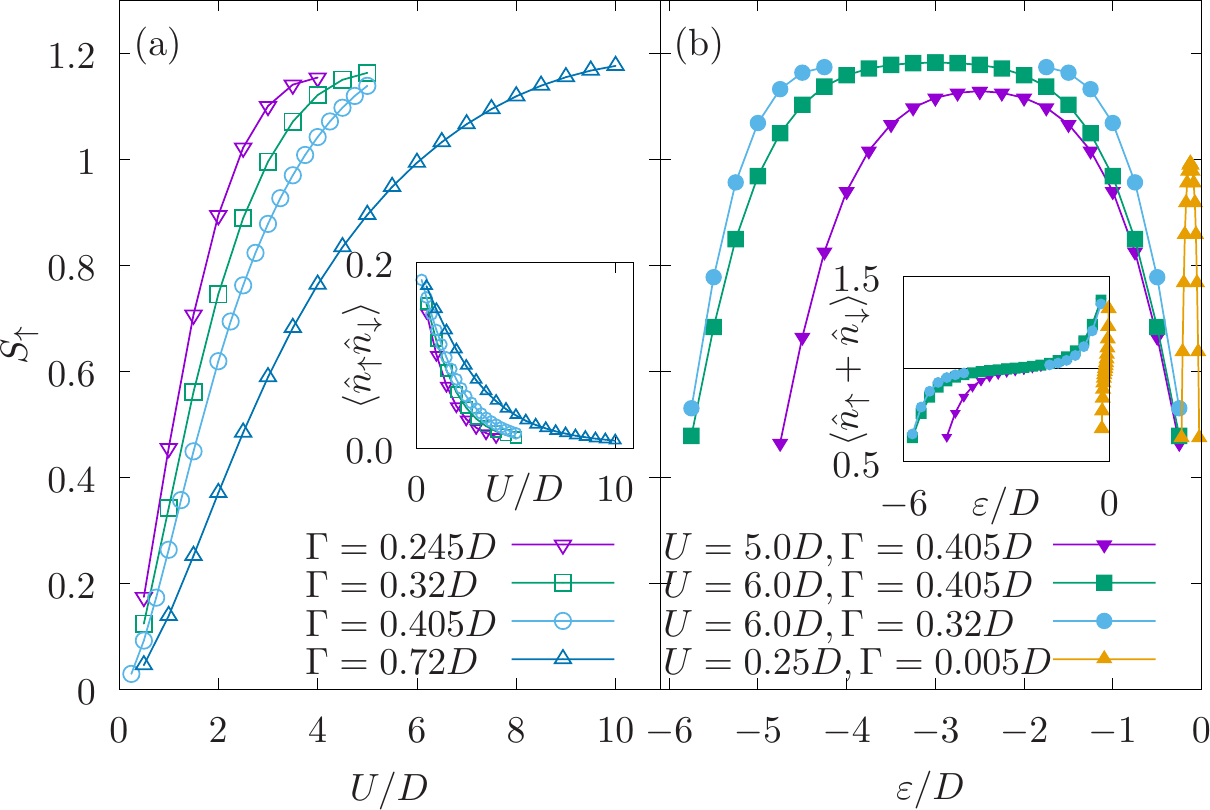}
	\caption{Spin entanglement entropy $S_\uparrow$ for a variety of the model parameters. (a) $S_\uparrow$ vs. $U$ for an electron-hole symmetric situation $\varepsilon=-U/2$. The inset shows the impurity level double occupancy probability. (b) $S_\uparrow$ vs. $\varepsilon$  for fixed local interaction as indicated in the figure. The entropy is symmetric under the transformation $\varepsilon\to -U-\varepsilon$ due to the electron-hole symmetry of the electron bath. The inset shows the level occupancy.}
\label{fig:params}
\end{figure}

Figure \ref{fig:params} presents the spin entanglement entropy $S_\uparrow$ for a variety of model parameters. $S_\uparrow$ decreases monotonically as the system is driven, by the model parameters, to a less correlated ground state, i.e. increasing the impurity-electron bath coupling, decreasing the Coulomb repulsion $U$ or shifting the level energy away from the electron-hole symmetric situation. In Fig. \ref{fig:params}a) the system is in an electron-hole symmetric regime with $\varepsilon=-U/2$ and the average impurity level occupancy $n_f$ is $1$. Increasing $U/\Gamma$ leads to a reduction in the average double occupancy [see inset to  Fig. \ref{fig:params}a)] which signals an increase in the correlations between opposite spin projection electrons at the impurity. In Fig. \ref{fig:params}b) the local interaction $U$ and the hybridization $\Gamma$ are fixed and the impurity level energy is shifted. The larger values of $S_\uparrow$ are obtained in the electron-hole-symmetric condition. As $n_f$ decreases from $1$  the interaction is less effective creating correlations between opposite spin projection electrons. A decreasing $n_f<1$ implies a larger probability of finding the system with an empty impurity level in which the interaction is not active.   The same argument is valid for $n_f>1$ due to the electron-hole symmetry \footnote{The same qualitative behavior of the electron-electron correlations and of the spin entanglement in the ground state wavefunction can be observed in the $L=1$ case [including a single site in the conduction band of Eq. (\ref{eq:tbchain})] which can be solved analytically for the entanglement entropy in the ground state.} .

To calculate the quasiparticle mass enhancement we define the zero-temperature spin susceptibility \cite{wilson1975kondo}.
\begin{equation}
	\chi = \left.\frac{d m_f}{d h}\right|_{h\to 0},
\end{equation}
which measures the change in the spin polarization of the impurity in the ground state $m_f=\left\langle(\hat{n}_{\uparrow}-\hat{n}_{\downarrow})\right\rangle/2$ when a Zeeman energy splitting $2 h= g\mu_B B$ is applied at the impurity.
In the numerical calculations presented below a small enough energy splitting $\delta h$ is applied, such that the response is linear~\footnote{An energy shift $\delta h=0.0001D$ proved to be appropriate in the whole parameter regime studied.}.
In the Kondo regime the low energy properties of the system are universal functions when properly scaled by the Kondo energy $\varepsilon_K\propto 1/\chi$ \cite{wilson1975kondo} and the quasiparticle mass enhancement can be estimated as $Z\sim(\Gamma \chi)^{-1}$ \cite{hewson1997kondo}. 
\begin{figure}[tb]\center
\includegraphics[width=\columnwidth,angle=0,keepaspectratio=true]{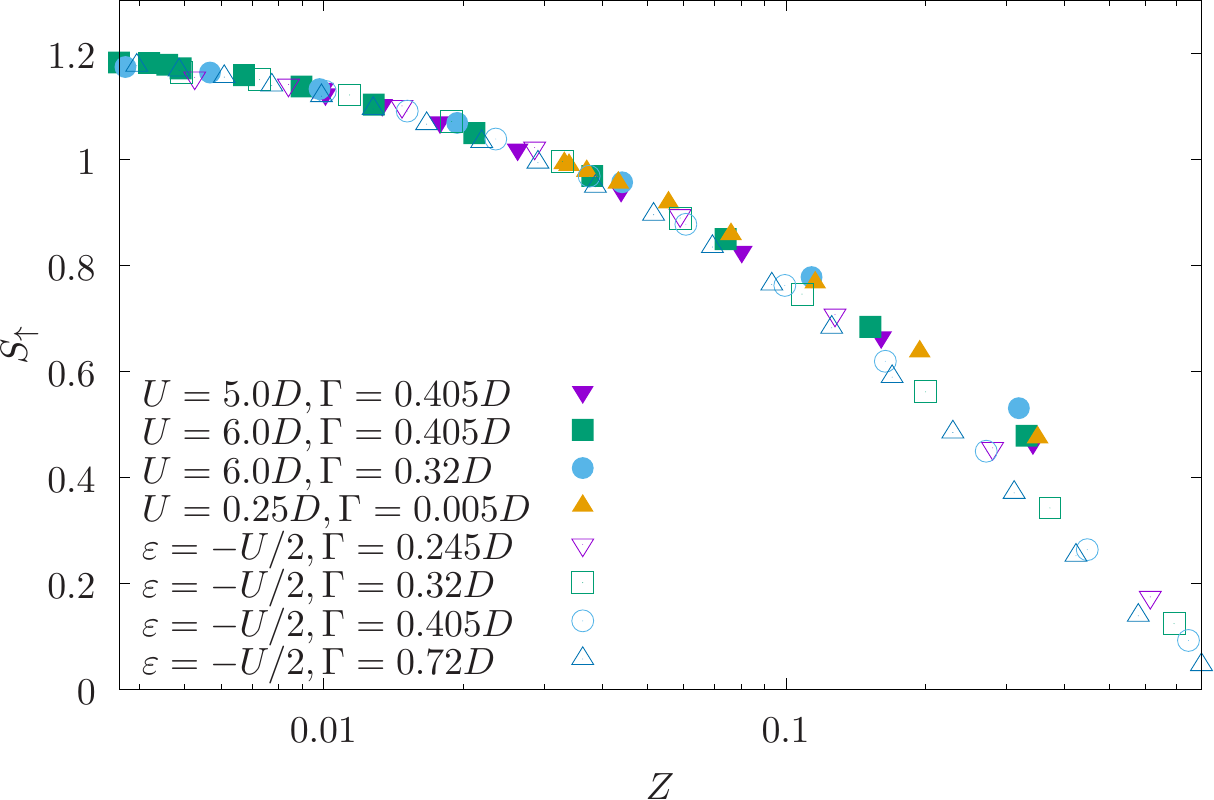}
\caption{Spin entanglement $S_\uparrow$ as a function of $Z$ for a variety of model parameters. In the low $Z$ regime, the data fall into a universal curve.  
}
\label{fig:Univ}
\end{figure}

Figure \ref{fig:Univ} presents the spin entanglement entropy as a function of the quasiparticle mass enhancement. For values of $Z\lesssim 0.1$ the data from Fig. \ref{fig:params} fall into a single curve as expected from the large-$N$ analysis. 
This universal and monotonic behavior indicates that the spin entanglement entropy is uniquely determined by the quasiparticle mass enhancement $Z$. 

There are several important differences between the numerical results for $N=2$ and the large-$N$ limit. In the latter $S_\uparrow\leq 1$ while in the $N=2$ case it shows values larger than $1$. This is due to the $N\to\infty$ limit and already including terms to order $1/N$  leads to $S_\uparrow> 1$ in the strongly correlated regime \footnote{We checked this numerically obtaining the wavefunction to order $1/N$ numerically for finite systems.}.

\begin{figure}[tb]\center
\includegraphics[width=\columnwidth,angle=0,keepaspectratio=true]{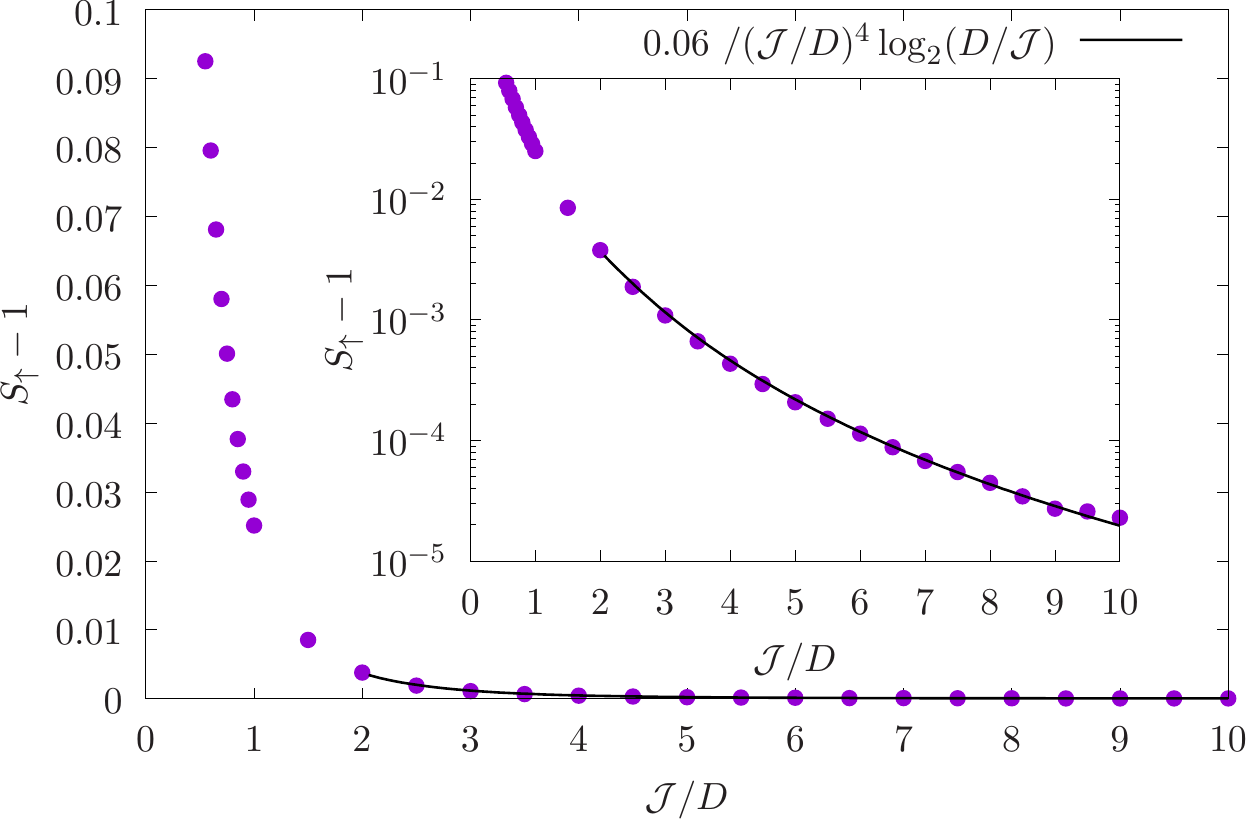}
	\caption{Spin entanglement entropy $S_\uparrow$ as a function of the exchange coupling $J$ in the Kondo model calculated using DMRG. The solid line is a fit using the functional form expected from perturbation theory in $D/\mathcal{J}$.}
\label{fig:entropJ}
\end{figure}

\paragraph{\textit{The $N=2$ Kondo model.\textemdash}} It is interesting to compare the spin entanglement entropy $S_\uparrow$ with the impurity-bath entanglement entropy $S_i$ to see whether they convey similar information. 
To that aim we focus on the $N=2$ case in the Kondo limit $\Gamma \ll |\varepsilon|, U$ in which we can ignore charge fluctuations at the impurity and only consider a magnetic exchange interaction $\mathcal{J}$ between a local magnetic moment in the impurity and the conduction bath $H_K=\mathcal{J} {\bm S}_f\cdot {\bm S}_1$, where ${\bm S}_\alpha=\frac{1}{2} \{c_{\alpha\uparrow}^\dagger,c_{\alpha\downarrow}^\dagger\}\cdot \bm{\sigma} \cdot \{c_{\alpha\uparrow},c_{\alpha\downarrow}\}^T$, and ${\bm{\sigma}}$ is the Pauli vector. This is the Kondo model which can be obtained from the Anderson model in second order perturbation theory on the impurity-bath coupling \cite{Schrieffer-Wolff}, and $\mathcal{J}$ is a function of the Anderson model parameters. In this model, $S_i$ is trivially 1 for any value of $\mathcal{J}>0$, as the impurity is in a maximally entangled state with the bath, while $S_\uparrow\geq 1$ depends on the value of $\mathcal{J}$ as it can be seen by numerical calculations or by perturbation theory in $D/\mathcal{J}$ (see Fig. \ref{fig:entropJ}) \cite{suppmat}. In the $\mathcal{J}/D \to \infty$ limit, the hopping terms can be neglected and the ground state is a spin singlet formed by a spin $1/2$ at the impurity and a spin $1/2$ at site $1$ of the tight binding chain. This readily leads to $S_\uparrow(\mathcal{J}\to \infty)=1$ and perturbation theory in $D/\mathcal{J}$ leads to a positive correction $\propto \frac{D^4}{\mathcal{J}^4} \log_2(\mathcal{J}/D)$. The numerical calculations show a monotonic increase in $S_\uparrow$ as $\mathcal{J}$ is decreased. 
These results for the Kondo model show that the spin entanglement conveys more information about interaction induced correlations than the impurity bath entanglement. 

In summary, we have found through analytical methods in the large-$N$ and large-$U$ limits and by numerical approaches in the finite $U$ and $N=2$ case, that the spin entanglement is univocally associated with the interaction induced quasiparticle mass enhancement in the Kondo regime. The quasiparticle mass enhancement can be obtained from spectroscopic transport measurements of the Kondo resonance \cite{PhysRevLett.80.2893,Madhavan567} which would allow obtaining the spin entanglement in the ground state wavefunction (see also Ref. \cite{PhysRevLett.120.146801}).

The quasiparticle mass enhancement plays also a crucial role when characterizing strong electronic correlations in heavy fermion materials and to assess the proximity to a Mott's metal-insulator transition, while DMFT establishes a connection between the physics of strongly correlated electron materials and quantum impurity problems \cite{georges1996dynamical,georges2015beauty}. In DMFT the lattice problem is reduced to an impurity problem with a self-consistent electron bath, which in the case of the Hubbard model, is the Anderson impurity model. It would be of interest to exploit this connection to analyze if the quasiparticle mass enhancement in the Hubbard model can be interpreted as characterizing the interaction induced entanglement in the ground state wavefunction. 



\acknowledgments
This work was partially supported by ANPCyT PICT 2016-0204.
\bibliographystyle{apsrev4-1}
\bibliography{references}

\begin{thebibliography}{43}%
\makeatletter
\providecommand \@ifxundefined [1]{%
 \@ifx{#1\undefined}
}%
\providecommand \@ifnum [1]{%
 \ifnum #1\expandafter \@firstoftwo
 \else \expandafter \@secondoftwo
 \fi
}%
\providecommand \@ifx [1]{%
 \ifx #1\expandafter \@firstoftwo
 \else \expandafter \@secondoftwo
 \fi
}%
\providecommand \natexlab [1]{#1}%
\providecommand \enquote  [1]{``#1''}%
\providecommand \bibnamefont  [1]{#1}%
\providecommand \bibfnamefont [1]{#1}%
\providecommand \citenamefont [1]{#1}%
\providecommand \href@noop [0]{\@secondoftwo}%
\providecommand \href [0]{\begingroup \@sanitize@url \@href}%
\providecommand \@href[1]{\@@startlink{#1}\@@href}%
\providecommand \@@href[1]{\endgroup#1\@@endlink}%
\providecommand \@sanitize@url [0]{\catcode `\\12\catcode `\$12\catcode
  `\&12\catcode `\#12\catcode `\^12\catcode `\_12\catcode `\%12\relax}%
\providecommand \@@startlink[1]{}%
\providecommand \@@endlink[0]{}%
\providecommand \url  [0]{\begingroup\@sanitize@url \@url }%
\providecommand \@url [1]{\endgroup\@href {#1}{\urlprefix }}%
\providecommand \urlprefix  [0]{URL }%
\providecommand \Eprint [0]{\href }%
\providecommand \doibase [0]{http://dx.doi.org/}%
\providecommand \selectlanguage [0]{\@gobble}%
\providecommand \bibinfo  [0]{\@secondoftwo}%
\providecommand \bibfield  [0]{\@secondoftwo}%
\providecommand \translation [1]{[#1]}%
\providecommand \BibitemOpen [0]{}%
\providecommand \bibitemStop [0]{}%
\providecommand \bibitemNoStop [0]{.\EOS\space}%
\providecommand \EOS [0]{\spacefactor3000\relax}%
\providecommand \BibitemShut  [1]{\csname bibitem#1\endcsname}%
\let\auto@bib@innerbib\@empty
\bibitem [{\citenamefont {Osterloh}\ \emph {et~al.}(2002)\citenamefont
  {Osterloh}, \citenamefont {Amico}, \citenamefont {Falci},\ and\ \citenamefont
  {Fazio}}]{osterloh2002scaling}%
  \BibitemOpen
  \bibfield  {author} {\bibinfo {author} {\bibfnamefont {A.}~\bibnamefont
  {Osterloh}}, \bibinfo {author} {\bibfnamefont {L.}~\bibnamefont {Amico}},
  \bibinfo {author} {\bibfnamefont {G.}~\bibnamefont {Falci}}, \ and\ \bibinfo
  {author} {\bibfnamefont {R.}~\bibnamefont {Fazio}},\ }\href@noop {}
  {\bibfield  {journal} {\bibinfo  {journal} {Nature}\ }\textbf {\bibinfo
  {volume} {416}},\ \bibinfo {pages} {608} (\bibinfo {year}
  {2002})}\BibitemShut {NoStop}%
\bibitem [{\citenamefont {Osborne}\ and\ \citenamefont
  {Nielsen}(2002)}]{osborne2002entanglement}%
  \BibitemOpen
  \bibfield  {author} {\bibinfo {author} {\bibfnamefont {T.~J.}\ \bibnamefont
  {Osborne}}\ and\ \bibinfo {author} {\bibfnamefont {M.~A.}\ \bibnamefont
  {Nielsen}},\ }\href@noop {} {\bibfield  {journal} {\bibinfo  {journal}
  {Physical Review A}\ }\textbf {\bibinfo {volume} {66}},\ \bibinfo {pages}
  {032110} (\bibinfo {year} {2002})}\BibitemShut {NoStop}%
\bibitem [{\citenamefont {Gu}\ \emph {et~al.}(2004)\citenamefont {Gu},
  \citenamefont {Deng}, \citenamefont {Li},\ and\ \citenamefont
  {Lin}}]{gu2004entanglement}%
  \BibitemOpen
  \bibfield  {author} {\bibinfo {author} {\bibfnamefont {S.-J.}\ \bibnamefont
  {Gu}}, \bibinfo {author} {\bibfnamefont {S.-S.}\ \bibnamefont {Deng}},
  \bibinfo {author} {\bibfnamefont {Y.-Q.}\ \bibnamefont {Li}}, \ and\ \bibinfo
  {author} {\bibfnamefont {H.-Q.}\ \bibnamefont {Lin}},\ }\href@noop {}
  {\bibfield  {journal} {\bibinfo  {journal} {Physical review letters}\
  }\textbf {\bibinfo {volume} {93}},\ \bibinfo {pages} {086402} (\bibinfo
  {year} {2004})}\BibitemShut {NoStop}%
\bibitem [{\citenamefont {Bayat}\ \emph {et~al.}(2014)\citenamefont {Bayat},
  \citenamefont {Johannesson}, \citenamefont {Bose},\ and\ \citenamefont
  {Sodano}}]{bayat2014order}%
  \BibitemOpen
  \bibfield  {author} {\bibinfo {author} {\bibfnamefont {A.}~\bibnamefont
  {Bayat}}, \bibinfo {author} {\bibfnamefont {H.}~\bibnamefont {Johannesson}},
  \bibinfo {author} {\bibfnamefont {S.}~\bibnamefont {Bose}}, \ and\ \bibinfo
  {author} {\bibfnamefont {P.}~\bibnamefont {Sodano}},\ }\href@noop {}
  {\bibfield  {journal} {\bibinfo  {journal} {Nature communications}\ }\textbf
  {\bibinfo {volume} {5}},\ \bibinfo {pages} {3784} (\bibinfo {year}
  {2014})}\BibitemShut {NoStop}%
\bibitem [{\citenamefont {Deutsch}(1991)}]{deutsch1991quantum}%
  \BibitemOpen
  \bibfield  {author} {\bibinfo {author} {\bibfnamefont {J.~M.}\ \bibnamefont
  {Deutsch}},\ }\href@noop {} {\bibfield  {journal} {\bibinfo  {journal}
  {Physical Review A}\ }\textbf {\bibinfo {volume} {43}},\ \bibinfo {pages}
  {2046} (\bibinfo {year} {1991})}\BibitemShut {NoStop}%
\bibitem [{\citenamefont {Srednicki}(1994)}]{srednicki1994chaos}%
  \BibitemOpen
  \bibfield  {author} {\bibinfo {author} {\bibfnamefont {M.}~\bibnamefont
  {Srednicki}},\ }\href@noop {} {\bibfield  {journal} {\bibinfo  {journal}
  {Physical Review E}\ }\textbf {\bibinfo {volume} {50}},\ \bibinfo {pages}
  {888} (\bibinfo {year} {1994})}\BibitemShut {NoStop}%
\bibitem [{\citenamefont {Rigol}\ \emph {et~al.}(2008)\citenamefont {Rigol},
  \citenamefont {Dunjko},\ and\ \citenamefont
  {Olshanii}}]{rigol2008thermalization}%
  \BibitemOpen
  \bibfield  {author} {\bibinfo {author} {\bibfnamefont {M.}~\bibnamefont
  {Rigol}}, \bibinfo {author} {\bibfnamefont {V.}~\bibnamefont {Dunjko}}, \
  and\ \bibinfo {author} {\bibfnamefont {M.}~\bibnamefont {Olshanii}},\
  }\href@noop {} {\bibfield  {journal} {\bibinfo  {journal} {Nature}\ }\textbf
  {\bibinfo {volume} {452}},\ \bibinfo {pages} {854} (\bibinfo {year}
  {2008})}\BibitemShut {NoStop}%
\bibitem [{\citenamefont {Islam}\ \emph {et~al.}(2015)\citenamefont {Islam},
  \citenamefont {Ma}, \citenamefont {Preiss}, \citenamefont {Tai},
  \citenamefont {Lukin}, \citenamefont {Rispoli},\ and\ \citenamefont
  {Greiner}}]{islam2015measuring}%
  \BibitemOpen
  \bibfield  {author} {\bibinfo {author} {\bibfnamefont {R.}~\bibnamefont
  {Islam}}, \bibinfo {author} {\bibfnamefont {R.}~\bibnamefont {Ma}}, \bibinfo
  {author} {\bibfnamefont {P.~M.}\ \bibnamefont {Preiss}}, \bibinfo {author}
  {\bibfnamefont {M.~E.}\ \bibnamefont {Tai}}, \bibinfo {author} {\bibfnamefont
  {A.}~\bibnamefont {Lukin}}, \bibinfo {author} {\bibfnamefont
  {M.}~\bibnamefont {Rispoli}}, \ and\ \bibinfo {author} {\bibfnamefont
  {M.}~\bibnamefont {Greiner}},\ }\href@noop {} {\bibfield  {journal} {\bibinfo
   {journal} {Nature}\ }\textbf {\bibinfo {volume} {528}},\ \bibinfo {pages}
  {77} (\bibinfo {year} {2015})}\BibitemShut {NoStop}%
\bibitem [{\citenamefont {Kaufman}\ \emph {et~al.}(2016)\citenamefont
  {Kaufman}, \citenamefont {Tai}, \citenamefont {Lukin}, \citenamefont
  {Rispoli}, \citenamefont {Schittko}, \citenamefont {Preiss},\ and\
  \citenamefont {Greiner}}]{kaufman2016quantum}%
  \BibitemOpen
  \bibfield  {author} {\bibinfo {author} {\bibfnamefont {A.~M.}\ \bibnamefont
  {Kaufman}}, \bibinfo {author} {\bibfnamefont {M.~E.}\ \bibnamefont {Tai}},
  \bibinfo {author} {\bibfnamefont {A.}~\bibnamefont {Lukin}}, \bibinfo
  {author} {\bibfnamefont {M.}~\bibnamefont {Rispoli}}, \bibinfo {author}
  {\bibfnamefont {R.}~\bibnamefont {Schittko}}, \bibinfo {author}
  {\bibfnamefont {P.~M.}\ \bibnamefont {Preiss}}, \ and\ \bibinfo {author}
  {\bibfnamefont {M.}~\bibnamefont {Greiner}},\ }\href@noop {} {\bibfield
  {journal} {\bibinfo  {journal} {Science}\ }\textbf {\bibinfo {volume}
  {353}},\ \bibinfo {pages} {794} (\bibinfo {year} {2016})}\BibitemShut
  {NoStop}%
\bibitem [{\citenamefont {Lukin}\ \emph {et~al.}(2019)\citenamefont {Lukin},
  \citenamefont {Rispoli}, \citenamefont {Schittko}, \citenamefont {Tai},
  \citenamefont {Kaufman}, \citenamefont {Choi}, \citenamefont {Khemani},
  \citenamefont {L{\'e}onard},\ and\ \citenamefont
  {Greiner}}]{lukin2019probing}%
  \BibitemOpen
  \bibfield  {author} {\bibinfo {author} {\bibfnamefont {A.}~\bibnamefont
  {Lukin}}, \bibinfo {author} {\bibfnamefont {M.}~\bibnamefont {Rispoli}},
  \bibinfo {author} {\bibfnamefont {R.}~\bibnamefont {Schittko}}, \bibinfo
  {author} {\bibfnamefont {M.~E.}\ \bibnamefont {Tai}}, \bibinfo {author}
  {\bibfnamefont {A.~M.}\ \bibnamefont {Kaufman}}, \bibinfo {author}
  {\bibfnamefont {S.}~\bibnamefont {Choi}}, \bibinfo {author} {\bibfnamefont
  {V.}~\bibnamefont {Khemani}}, \bibinfo {author} {\bibfnamefont
  {J.}~\bibnamefont {L{\'e}onard}}, \ and\ \bibinfo {author} {\bibfnamefont
  {M.}~\bibnamefont {Greiner}},\ }\href@noop {} {\bibfield  {journal} {\bibinfo
   {journal} {Science}\ }\textbf {\bibinfo {volume} {364}},\ \bibinfo {pages}
  {256} (\bibinfo {year} {2019})}\BibitemShut {NoStop}%
\bibitem [{\citenamefont {Hewson}(1997)}]{hewson1997kondo}%
  \BibitemOpen
  \bibfield  {author} {\bibinfo {author} {\bibfnamefont {A.~C.}\ \bibnamefont
  {Hewson}},\ }\href@noop {} {\emph {\bibinfo {title} {The Kondo problem to
  heavy fermions}}},\ Vol.~\bibinfo {volume} {2}\ (\bibinfo  {publisher}
  {Cambridge university press},\ \bibinfo {year} {1997})\BibitemShut {NoStop}%
\bibitem [{\citenamefont {Wilson}(1975)}]{wilson1975kondo}%
  \BibitemOpen
  \bibfield  {author} {\bibinfo {author} {\bibfnamefont {K.~G.}\ \bibnamefont
  {Wilson}},\ }\href {\doibase 10.1103/RevModPhys.47.773} {\bibfield  {journal}
  {\bibinfo  {journal} {Rev. Mod. Phys.}\ }\textbf {\bibinfo {volume} {47}},\
  \bibinfo {pages} {773} (\bibinfo {year} {1975})}\BibitemShut {NoStop}%
\bibitem [{\citenamefont {Georges}(2016)}]{georges2015beauty}%
  \BibitemOpen
  \bibfield  {author} {\bibinfo {author} {\bibfnamefont {A.}~\bibnamefont
  {Georges}},\ }\href {\doibase http://dx.doi.org/10.1016/j.crhy.2015.12.005}
  {\bibfield  {journal} {\bibinfo  {journal} {Comptes Rendus Physique}\
  }\textbf {\bibinfo {volume} {17}},\ \bibinfo {pages} {430 } (\bibinfo {year}
  {2016})}\BibitemShut {NoStop}%
\bibitem [{\citenamefont {Costi}(2001)}]{costi2001magnetotransport}%
  \BibitemOpen
  \bibfield  {author} {\bibinfo {author} {\bibfnamefont {T.~A.}\ \bibnamefont
  {Costi}},\ }\href@noop {} {\bibfield  {journal} {\bibinfo  {journal}
  {Physical Review B}\ }\textbf {\bibinfo {volume} {64}},\ \bibinfo {pages}
  {241310(R)} (\bibinfo {year} {2001})}\BibitemShut {NoStop}%
\bibitem [{\citenamefont {Cornaglia}\ and\ \citenamefont
  {Balseiro}(2003)}]{cornaglia2003transport}%
  \BibitemOpen
  \bibfield  {author} {\bibinfo {author} {\bibfnamefont {P.~S.}\ \bibnamefont
  {Cornaglia}}\ and\ \bibinfo {author} {\bibfnamefont {C.~A.}\ \bibnamefont
  {Balseiro}},\ }\href@noop {} {\bibfield  {journal} {\bibinfo  {journal}
  {Physical review letters}\ }\textbf {\bibinfo {volume} {90}},\ \bibinfo
  {pages} {216801} (\bibinfo {year} {2003})}\BibitemShut {NoStop}%
\bibitem [{\citenamefont {Georges}\ \emph {et~al.}(1996)\citenamefont
  {Georges}, \citenamefont {Kotliar}, \citenamefont {Krauth},\ and\
  \citenamefont {Rozenberg}}]{georges1996dynamical}%
  \BibitemOpen
  \bibfield  {author} {\bibinfo {author} {\bibfnamefont {A.}~\bibnamefont
  {Georges}}, \bibinfo {author} {\bibfnamefont {G.}~\bibnamefont {Kotliar}},
  \bibinfo {author} {\bibfnamefont {W.}~\bibnamefont {Krauth}}, \ and\ \bibinfo
  {author} {\bibfnamefont {M.~J.}\ \bibnamefont {Rozenberg}},\ }\href {\doibase
  10.1103/RevModPhys.68.13} {\bibfield  {journal} {\bibinfo  {journal} {Rev.
  Mod. Phys.}\ }\textbf {\bibinfo {volume} {68}},\ \bibinfo {pages} {13}
  (\bibinfo {year} {1996})}\BibitemShut {NoStop}%
\bibitem [{\citenamefont {Nozieres}(1974)}]{nozieres1974fermi}%
  \BibitemOpen
  \bibfield  {author} {\bibinfo {author} {\bibfnamefont {P.}~\bibnamefont
  {Nozieres}},\ }\href@noop {} {\bibfield  {journal} {\bibinfo  {journal}
  {Journal of Low Temperature Physics}\ }\textbf {\bibinfo {volume} {17}},\
  \bibinfo {pages} {31} (\bibinfo {year} {1974})}\BibitemShut {NoStop}%
\bibitem [{\citenamefont {Fulde}(2012)}]{fulde2012correlated}%
  \BibitemOpen
  \bibfield  {author} {\bibinfo {author} {\bibfnamefont {P.}~\bibnamefont
  {Fulde}},\ }\href@noop {} {\emph {\bibinfo {title} {Correlated electrons in
  quantum matter}}}\ (\bibinfo  {publisher} {World Scientific},\ \bibinfo
  {year} {2012})\BibitemShut {NoStop}%
\bibitem [{\citenamefont {Imada}\ \emph {et~al.}(1998)\citenamefont {Imada},
  \citenamefont {Fujimori},\ and\ \citenamefont {Tokura}}]{masa1998mit}%
  \BibitemOpen
  \bibfield  {author} {\bibinfo {author} {\bibfnamefont {M.}~\bibnamefont
  {Imada}}, \bibinfo {author} {\bibfnamefont {A.}~\bibnamefont {Fujimori}}, \
  and\ \bibinfo {author} {\bibfnamefont {Y.}~\bibnamefont {Tokura}},\ }\href
  {\doibase 10.1103/RevModPhys.70.1039} {\bibfield  {journal} {\bibinfo
  {journal} {Rev. Mod. Phys.}\ }\textbf {\bibinfo {volume} {70}},\ \bibinfo
  {pages} {1039} (\bibinfo {year} {1998})}\BibitemShut {NoStop}%
\bibitem [{\citenamefont {Borzenets}\ \emph {et~al.}(2020)\citenamefont
  {Borzenets}, \citenamefont {Shim}, \citenamefont {Chen}, \citenamefont
  {Ludwig}, \citenamefont {Wieck}, \citenamefont {Tarucha}, \citenamefont
  {Sim},\ and\ \citenamefont {Yamamoto}}]{borzenets2020observation}%
  \BibitemOpen
  \bibfield  {author} {\bibinfo {author} {\bibfnamefont {I.~V.}\ \bibnamefont
  {Borzenets}}, \bibinfo {author} {\bibfnamefont {J.}~\bibnamefont {Shim}},
  \bibinfo {author} {\bibfnamefont {J.~C.}\ \bibnamefont {Chen}}, \bibinfo
  {author} {\bibfnamefont {A.}~\bibnamefont {Ludwig}}, \bibinfo {author}
  {\bibfnamefont {A.~D.}\ \bibnamefont {Wieck}}, \bibinfo {author}
  {\bibfnamefont {S.}~\bibnamefont {Tarucha}}, \bibinfo {author} {\bibfnamefont
  {H.-S.}\ \bibnamefont {Sim}}, \ and\ \bibinfo {author} {\bibfnamefont
  {M.}~\bibnamefont {Yamamoto}},\ }\href@noop {} {\bibfield  {journal}
  {\bibinfo  {journal} {Nature}\ }\textbf {\bibinfo {volume} {579}},\ \bibinfo
  {pages} {210} (\bibinfo {year} {2020})}\BibitemShut {NoStop}%
\bibitem [{\citenamefont {S{\o}rensen}\ \emph {et~al.}(2007)\citenamefont
  {S{\o}rensen}, \citenamefont {Chang}, \citenamefont {Laflorencie},\ and\
  \citenamefont {Affleck}}]{sorensen2007impurity}%
  \BibitemOpen
  \bibfield  {author} {\bibinfo {author} {\bibfnamefont {E.~S.}\ \bibnamefont
  {S{\o}rensen}}, \bibinfo {author} {\bibfnamefont {M.-S.}\ \bibnamefont
  {Chang}}, \bibinfo {author} {\bibfnamefont {N.}~\bibnamefont {Laflorencie}},
  \ and\ \bibinfo {author} {\bibfnamefont {I.}~\bibnamefont {Affleck}},\
  }\href@noop {} {\bibfield  {journal} {\bibinfo  {journal} {Journal of
  Statistical Mechanics: Theory and Experiment}\ }\textbf {\bibinfo {volume}
  {2007}},\ \bibinfo {pages} {L01001} (\bibinfo {year} {2007})}\BibitemShut
  {NoStop}%
\bibitem [{\citenamefont {Bayat}\ \emph {et~al.}(2010)\citenamefont {Bayat},
  \citenamefont {Sodano},\ and\ \citenamefont {Bose}}]{bayat2010negativity}%
  \BibitemOpen
  \bibfield  {author} {\bibinfo {author} {\bibfnamefont {A.}~\bibnamefont
  {Bayat}}, \bibinfo {author} {\bibfnamefont {P.}~\bibnamefont {Sodano}}, \
  and\ \bibinfo {author} {\bibfnamefont {S.}~\bibnamefont {Bose}},\ }\href
  {\doibase 10.1103/PhysRevB.81.064429} {\bibfield  {journal} {\bibinfo
  {journal} {Phys. Rev. B}\ }\textbf {\bibinfo {volume} {81}},\ \bibinfo
  {pages} {064429} (\bibinfo {year} {2010})}\BibitemShut {NoStop}%
\bibitem [{\citenamefont {Bayat}\ \emph {et~al.}(2012)\citenamefont {Bayat},
  \citenamefont {Bose}, \citenamefont {Sodano},\ and\ \citenamefont
  {Johannesson}}]{bayat2012entanglement}%
  \BibitemOpen
  \bibfield  {author} {\bibinfo {author} {\bibfnamefont {A.}~\bibnamefont
  {Bayat}}, \bibinfo {author} {\bibfnamefont {S.}~\bibnamefont {Bose}},
  \bibinfo {author} {\bibfnamefont {P.}~\bibnamefont {Sodano}}, \ and\ \bibinfo
  {author} {\bibfnamefont {H.}~\bibnamefont {Johannesson}},\ }\href {\doibase
  10.1103/PhysRevLett.109.066403} {\bibfield  {journal} {\bibinfo  {journal}
  {Phys. Rev. Lett.}\ }\textbf {\bibinfo {volume} {109}},\ \bibinfo {pages}
  {066403} (\bibinfo {year} {2012})}\BibitemShut {NoStop}%
\bibitem [{\citenamefont {Alkurtass}\ \emph {et~al.}(2016)\citenamefont
  {Alkurtass}, \citenamefont {Bayat}, \citenamefont {Affleck}, \citenamefont
  {Bose}, \citenamefont {Johannesson}, \citenamefont {Sodano}, \citenamefont
  {S\o{}rensen},\ and\ \citenamefont {Le~Hur}}]{alkurtass2016entanglement}%
  \BibitemOpen
  \bibfield  {author} {\bibinfo {author} {\bibfnamefont {B.}~\bibnamefont
  {Alkurtass}}, \bibinfo {author} {\bibfnamefont {A.}~\bibnamefont {Bayat}},
  \bibinfo {author} {\bibfnamefont {I.}~\bibnamefont {Affleck}}, \bibinfo
  {author} {\bibfnamefont {S.}~\bibnamefont {Bose}}, \bibinfo {author}
  {\bibfnamefont {H.}~\bibnamefont {Johannesson}}, \bibinfo {author}
  {\bibfnamefont {P.}~\bibnamefont {Sodano}}, \bibinfo {author} {\bibfnamefont
  {E.~S.}\ \bibnamefont {S\o{}rensen}}, \ and\ \bibinfo {author} {\bibfnamefont
  {K.}~\bibnamefont {Le~Hur}},\ }\href {\doibase 10.1103/PhysRevB.93.081106}
  {\bibfield  {journal} {\bibinfo  {journal} {Phys. Rev. B}\ }\textbf {\bibinfo
  {volume} {93}},\ \bibinfo {pages} {081106(R)} (\bibinfo {year}
  {2016})}\BibitemShut {NoStop}%
\bibitem [{\citenamefont {Laflorencie}(2016)}]{Laflorencie20161}%
  \BibitemOpen
  \bibfield  {author} {\bibinfo {author} {\bibfnamefont {N.}~\bibnamefont
  {Laflorencie}},\ }\href {\doibase
  http://dx.doi.org/10.1016/j.physrep.2016.06.008} {\bibfield  {journal}
  {\bibinfo  {journal} {Physics Reports}\ }\textbf {\bibinfo {volume} {646}},\
  \bibinfo {pages} {1 } (\bibinfo {year} {2016})},\ \bibinfo {note} {quantum
  entanglement in condensed matter systems}\BibitemShut {NoStop}%
\bibitem [{\citenamefont {Bayat}(2017)}]{bayat2017scaling}%
  \BibitemOpen
  \bibfield  {author} {\bibinfo {author} {\bibfnamefont {A.}~\bibnamefont
  {Bayat}},\ }\href {\doibase 10.1103/PhysRevLett.118.036102} {\bibfield
  {journal} {\bibinfo  {journal} {Phys. Rev. Lett.}\ }\textbf {\bibinfo
  {volume} {118}},\ \bibinfo {pages} {036102} (\bibinfo {year}
  {2017})}\BibitemShut {NoStop}%
\bibitem [{\citenamefont {Yang}\ and\ \citenamefont
  {Feiguin}(2017)}]{feiguin2017}%
  \BibitemOpen
  \bibfield  {author} {\bibinfo {author} {\bibfnamefont {C.}~\bibnamefont
  {Yang}}\ and\ \bibinfo {author} {\bibfnamefont {A.~E.}\ \bibnamefont
  {Feiguin}},\ }\href {\doibase 10.1103/PhysRevB.95.115106} {\bibfield
  {journal} {\bibinfo  {journal} {Phys. Rev. B}\ }\textbf {\bibinfo {volume}
  {95}},\ \bibinfo {pages} {115106} (\bibinfo {year} {2017})}\BibitemShut
  {NoStop}%
\bibitem [{\citenamefont {Wagner}\ \emph {et~al.}(2018)\citenamefont {Wagner},
  \citenamefont {Chowdhury}, \citenamefont {Pixley},\ and\ \citenamefont
  {Ingersent}}]{PhysRevLett.121.147602}%
  \BibitemOpen
  \bibfield  {author} {\bibinfo {author} {\bibfnamefont {C.}~\bibnamefont
  {Wagner}}, \bibinfo {author} {\bibfnamefont {T.}~\bibnamefont {Chowdhury}},
  \bibinfo {author} {\bibfnamefont {J.~H.}\ \bibnamefont {Pixley}}, \ and\
  \bibinfo {author} {\bibfnamefont {K.}~\bibnamefont {Ingersent}},\ }\href
  {\doibase 10.1103/PhysRevLett.121.147602} {\bibfield  {journal} {\bibinfo
  {journal} {Phys. Rev. Lett.}\ }\textbf {\bibinfo {volume} {121}},\ \bibinfo
  {pages} {147602} (\bibinfo {year} {2018})}\BibitemShut {NoStop}%
\bibitem [{\citenamefont {Bickers}(1987)}]{bickers1987review}%
  \BibitemOpen
  \bibfield  {author} {\bibinfo {author} {\bibfnamefont {N.}~\bibnamefont
  {Bickers}},\ }\href@noop {} {\bibfield  {journal} {\bibinfo  {journal}
  {Reviews of modern physics}\ }\textbf {\bibinfo {volume} {59}},\ \bibinfo
  {pages} {845} (\bibinfo {year} {1987})}\BibitemShut {NoStop}%
\bibitem [{\citenamefont {Varma}\ and\ \citenamefont
  {Yafet}(1976)}]{varma1975}%
  \BibitemOpen
  \bibfield  {author} {\bibinfo {author} {\bibfnamefont {C.~M.}\ \bibnamefont
  {Varma}}\ and\ \bibinfo {author} {\bibfnamefont {Y.}~\bibnamefont {Yafet}},\
  }\href {\doibase 10.1103/PhysRevB.13.2950} {\bibfield  {journal} {\bibinfo
  {journal} {Phys. Rev. B}\ }\textbf {\bibinfo {volume} {13}},\ \bibinfo
  {pages} {2950} (\bibinfo {year} {1976})}\BibitemShut {NoStop}%
\bibitem [{sup()}]{suppmat}%
  \BibitemOpen
  \href@noop {} {}\bibinfo {note} {See supplemental material at
  URL.}\BibitemShut {Stop}%
\bibitem [{\citenamefont {Peschel}\ \emph {et~al.}(1999)\citenamefont
  {Peschel}, \citenamefont {Want}, \citenamefont {Kaulke},\ and\ \citenamefont
  {Hallberg}}]{peschel1999density}%
  \BibitemOpen
  \bibfield  {author} {\bibinfo {author} {\bibfnamefont {I.}~\bibnamefont
  {Peschel}}, \bibinfo {author} {\bibfnamefont {X.}~\bibnamefont {Want}},
  \bibinfo {author} {\bibfnamefont {M.}~\bibnamefont {Kaulke}}, \ and\ \bibinfo
  {author} {\bibfnamefont {K.}~\bibnamefont {Hallberg}},\ }in\ \href@noop {}
  {\emph {\bibinfo {booktitle} {Density-matrix renormalization, a new numerical
  method in physics}}},\ Vol.\ \bibinfo {volume} {528}\ (\bibinfo {year}
  {1999})\BibitemShut {NoStop}%
\bibitem [{\citenamefont {Hallberg}(2006)}]{hallberg2006new}%
  \BibitemOpen
  \bibfield  {author} {\bibinfo {author} {\bibfnamefont {K.~A.}\ \bibnamefont
  {Hallberg}},\ }\href@noop {} {\bibfield  {journal} {\bibinfo  {journal}
  {Advances in Physics}\ }\textbf {\bibinfo {volume} {55}},\ \bibinfo {pages}
  {477} (\bibinfo {year} {2006})}\BibitemShut {NoStop}%
\bibitem [{\citenamefont {Schollw{\"o}ck}(2005)}]{schollwock2005density}%
  \BibitemOpen
  \bibfield  {author} {\bibinfo {author} {\bibfnamefont {U.}~\bibnamefont
  {Schollw{\"o}ck}},\ }\href@noop {} {\bibfield  {journal} {\bibinfo  {journal}
  {Reviews of modern physics}\ }\textbf {\bibinfo {volume} {77}},\ \bibinfo
  {pages} {259} (\bibinfo {year} {2005})}\BibitemShut {NoStop}%
\bibitem [{\citenamefont {Thimm}\ \emph {et~al.}(1999)\citenamefont {Thimm},
  \citenamefont {Kroha},\ and\ \citenamefont {von Delft}}]{thimm1999kondo}%
  \BibitemOpen
  \bibfield  {author} {\bibinfo {author} {\bibfnamefont {W.~B.}\ \bibnamefont
  {Thimm}}, \bibinfo {author} {\bibfnamefont {J.}~\bibnamefont {Kroha}}, \ and\
  \bibinfo {author} {\bibfnamefont {J.}~\bibnamefont {von Delft}},\ }\href@noop
  {} {\bibfield  {journal} {\bibinfo  {journal} {Physical review letters}\
  }\textbf {\bibinfo {volume} {82}},\ \bibinfo {pages} {2143} (\bibinfo {year}
  {1999})}\BibitemShut {NoStop}%
\bibitem [{\citenamefont {Cornaglia}\ and\ \citenamefont
  {Balseiro}(2002)}]{cornaglia2002mesokondo}%
  \BibitemOpen
  \bibfield  {author} {\bibinfo {author} {\bibfnamefont {P.~S.}\ \bibnamefont
  {Cornaglia}}\ and\ \bibinfo {author} {\bibfnamefont {C.~A.}\ \bibnamefont
  {Balseiro}},\ }\href {\doibase 10.1103/PhysRevB.66.115303} {\bibfield
  {journal} {\bibinfo  {journal} {Phys. Rev. B}\ }\textbf {\bibinfo {volume}
  {66}},\ \bibinfo {pages} {115303} (\bibinfo {year} {2002})}\BibitemShut
  {NoStop}%
\bibitem [{Note1()}]{Note1}%
  \BibitemOpen
  \bibinfo {note} {The same qualitative behavior of the electron-electron
  correlations and of the spin entanglement in the ground state wavefunction
  can be observed in the $L=1$ case [including a single site in the conduction
  band of Eq. (\ref {eq:tbchain})] which can be solved analytically for the
  entanglement entropy in the ground state.}\BibitemShut {Stop}%
\bibitem [{Note2()}]{Note2}%
  \BibitemOpen
  \bibinfo {note} {An energy shift $\delta h=0.0001D$ proved to be appropriate
  in the whole parameter regime studied.}\BibitemShut {Stop}%
\bibitem [{Note3()}]{Note3}%
  \BibitemOpen
  \bibinfo {note} {We checked this numerically obtaining the wave function to
  order $1/N$ numerically for finite systems.}\BibitemShut {Stop}%
\bibitem [{\citenamefont {Schrieffer}\ and\ \citenamefont
  {Wolff}(1966)}]{Schrieffer-Wolff}%
  \BibitemOpen
  \bibfield  {author} {\bibinfo {author} {\bibfnamefont {J.~R.}\ \bibnamefont
  {Schrieffer}}\ and\ \bibinfo {author} {\bibfnamefont {P.~A.}\ \bibnamefont
  {Wolff}},\ }\href {\doibase 10.1103/PhysRev.149.491} {\bibfield  {journal}
  {\bibinfo  {journal} {Phys. Rev.}\ }\textbf {\bibinfo {volume} {149}},\
  \bibinfo {pages} {491} (\bibinfo {year} {1966})}\BibitemShut {NoStop}%
\bibitem [{\citenamefont {Li}\ \emph {et~al.}(1998)\citenamefont {Li},
  \citenamefont {Schneider}, \citenamefont {Berndt},\ and\ \citenamefont
  {Delley}}]{PhysRevLett.80.2893}%
  \BibitemOpen
  \bibfield  {author} {\bibinfo {author} {\bibfnamefont {J.}~\bibnamefont
  {Li}}, \bibinfo {author} {\bibfnamefont {W.-D.}\ \bibnamefont {Schneider}},
  \bibinfo {author} {\bibfnamefont {R.}~\bibnamefont {Berndt}}, \ and\ \bibinfo
  {author} {\bibfnamefont {B.}~\bibnamefont {Delley}},\ }\href {\doibase
  10.1103/PhysRevLett.80.2893} {\bibfield  {journal} {\bibinfo  {journal}
  {Phys. Rev. Lett.}\ }\textbf {\bibinfo {volume} {80}},\ \bibinfo {pages}
  {2893} (\bibinfo {year} {1998})}\BibitemShut {NoStop}%
\bibitem [{\citenamefont {Madhavan}\ \emph {et~al.}(1998)\citenamefont
  {Madhavan}, \citenamefont {Chen}, \citenamefont {Jamneala}, \citenamefont
  {Crommie},\ and\ \citenamefont {Wingreen}}]{Madhavan567}%
  \BibitemOpen
  \bibfield  {author} {\bibinfo {author} {\bibfnamefont {V.}~\bibnamefont
  {Madhavan}}, \bibinfo {author} {\bibfnamefont {W.}~\bibnamefont {Chen}},
  \bibinfo {author} {\bibfnamefont {T.}~\bibnamefont {Jamneala}}, \bibinfo
  {author} {\bibfnamefont {M.~F.}\ \bibnamefont {Crommie}}, \ and\ \bibinfo
  {author} {\bibfnamefont {N.~S.}\ \bibnamefont {Wingreen}},\ }\href {\doibase
  10.1126/science.280.5363.567} {\bibfield  {journal} {\bibinfo  {journal}
  {Science}\ }\textbf {\bibinfo {volume} {280}},\ \bibinfo {pages} {567}
  (\bibinfo {year} {1998})}\BibitemShut {NoStop}%
\bibitem [{\citenamefont {Yoo}\ \emph {et~al.}(2018)\citenamefont {Yoo},
  \citenamefont {Lee},\ and\ \citenamefont {Sim}}]{PhysRevLett.120.146801}%
  \BibitemOpen
  \bibfield  {author} {\bibinfo {author} {\bibfnamefont {G.}~\bibnamefont
  {Yoo}}, \bibinfo {author} {\bibfnamefont {S.~S.~B.}\ \bibnamefont {Lee}}, \
  and\ \bibinfo {author} {\bibfnamefont {H.~S.}\ \bibnamefont {Sim}},\ }\href
  {\doibase 10.1103/PhysRevLett.120.146801} {\bibfield  {journal} {\bibinfo
  {journal} {Phys. Rev. Lett.}\ }\textbf {\bibinfo {volume} {120}},\ \bibinfo
  {pages} {146801} (\bibinfo {year} {2018})}\BibitemShut {NoStop}%
\end{thebibliography}%


\begin{thebibliography}{0}%
\makeatletter
\providecommand \@ifxundefined [1]{%
 \@ifx{#1\undefined}
}%
\providecommand \@ifnum [1]{%
 \ifnum #1\expandafter \@firstoftwo
 \else \expandafter \@secondoftwo
 \fi
}%
\providecommand \@ifx [1]{%
 \ifx #1\expandafter \@firstoftwo
 \else \expandafter \@secondoftwo
 \fi
}%
\providecommand \natexlab [1]{#1}%
\providecommand \enquote  [1]{``#1''}%
\providecommand \bibnamefont  [1]{#1}%
\providecommand \bibfnamefont [1]{#1}%
\providecommand \citenamefont [1]{#1}%
\providecommand \href@noop [0]{\@secondoftwo}%
\providecommand \href [0]{\begingroup \@sanitize@url \@href}%
\providecommand \@href[1]{\@@startlink{#1}\@@href}%
\providecommand \@@href[1]{\endgroup#1\@@endlink}%
\providecommand \@sanitize@url [0]{\catcode `\\12\catcode `\$12\catcode
  `\&12\catcode `\#12\catcode `\^12\catcode `\_12\catcode `\%12\relax}%
\providecommand \@@startlink[1]{}%
\providecommand \@@endlink[0]{}%
\providecommand \url  [0]{\begingroup\@sanitize@url \@url }%
\providecommand \@url [1]{\endgroup\@href {#1}{\urlprefix }}%
\providecommand \urlprefix  [0]{URL }%
\providecommand \Eprint [0]{\href }%
\providecommand \doibase [0]{http://dx.doi.org/}%
\providecommand \selectlanguage [0]{\@gobble}%
\providecommand \bibinfo  [0]{\@secondoftwo}%
\providecommand \bibfield  [0]{\@secondoftwo}%
\providecommand \translation [1]{[#1]}%
\providecommand \BibitemOpen [0]{}%
\providecommand \bibitemStop [0]{}%
\providecommand \bibitemNoStop [0]{.\EOS\space}%
\providecommand \EOS [0]{\spacefactor3000\relax}%
\providecommand \BibitemShut  [1]{\csname bibitem#1\endcsname}%
\let\auto@bib@innerbib\@empty
\end{thebibliography}%

\end{document}


\title{Supplemental material for: Quasiparticle mass enhancement as a measure of entanglement in the Kondo problem}

\author{Nayra A. \'Alvarez Pari}
\affiliation{Centro At{\'o}mico Bariloche and Instituto Balseiro, CNEA, 8400 Bariloche, Argentina}
\author{D. J.  Garc\'ia}
\affiliation{Centro At{\'o}mico Bariloche and Instituto Balseiro, CNEA, 8400 Bariloche, Argentina}
\affiliation{Consejo Nacional de Investigaciones Cient\'{\i}ficas y T\'ecnicas (CONICET), 8400 Bariloche, Argentina}
\author{Pablo S. Cornaglia}
\affiliation{Centro At{\'o}mico Bariloche and Instituto Balseiro, CNEA, 8400 Bariloche, Argentina}
\affiliation{Consejo Nacional de Investigaciones Cient\'{\i}ficas y T\'ecnicas (CONICET), 8400 Bariloche, Argentina}

\begin{abstract}
\end{abstract}

\maketitle
\section{Calculation of the Entanglement entropy in the $U\to\infty$ and $N\to \infty$ limits}
In these limits the ground state wave function is of the form (see main text for the notation) 
\begin{equation}
	|\Psi_{\text{GS}}\rangle=a_0\left( |F\rangle +\frac{1}{\sqrt{N}}\sum_{kj} b_k f_{j}^\dagger  c_{kj}|F\rangle\right)
\end{equation}
which can be written as
\begin{equation}
	|\Psi_{\text{GS}}\rangle=a_0 |F\rangle +\frac{a_0 b_0}{\sqrt{N}}\sum_{j}  f_{j}^\dagger  \tilde{c}_{j}|F\rangle
\end{equation}
where $\tilde{c}_j=\frac{1}{b_0}\sum_{\varepsilon_k \leq\varepsilon_F} b_k c_{kj}$ and $b_0^2=\sum_{\varepsilon_k \leq\varepsilon_F} b_k^2$ is a normalizing factor.

Separating positive and negative values of $j$ we have:
\begin{equation}
	|\Psi_{\text{GS}}\rangle=a_0 |F_{j\leq 0}\rangle\otimes |F_{j>0}\rangle +\frac{a_0 b_0}{\sqrt{N}}|F_{j\leq 0}\rangle\otimes\sum_{j>0} f_{j}^\dagger  \tilde{c}_{j}|F_{j>0}\rangle+\frac{a_0 b_0}{\sqrt{N}} \left( \sum_{j\leq 0} f_{j}^\dagger  \tilde{c}_{j}|F_{j\leq 0}\rangle \right)\otimes |F_{j> 0}\rangle
\end{equation}
where $|F_{j>0}\rangle=\prod_{\varepsilon_k\leq\varepsilon_F}\prod_{j>0}c_{kj}^\dagger|0\rangle$ and $|F_{j\leq 0}\rangle=\prod_{\varepsilon_k\leq\varepsilon_F}\prod_{j\leq 0}c_{kj}^\dagger|0\rangle$.

The parity of $N$ becomes irrelevant in the large-$N$ limit (the contribution of $j=0$ to the entropy vanishes) and we may assume $N$ to be an even number (i.e., no $j=0$ projection). We define the normalized states:
\begin{eqnarray}
  |\Psi_{j>0}\rangle = \sqrt{\frac{2}{N}}\sum_{j>0} f_{j}^\dagger  \tilde{c}_{j}|F_{j>0}\rangle,\\
  |\Psi_{j<0}\rangle = \sqrt{\frac{2}{N}}\sum_{j<0} f_{j}^\dagger  \tilde{c}_{j}|F_{j<0}\rangle,
\end{eqnarray}
which allow us to write the ground state wavefunction as:

\begin{equation}
	|\Psi_{\text{GS}}\rangle=a_0 |F_{j< 0}\rangle\otimes |F_{j>0}\rangle +\frac{a_0 b_0}{\sqrt{2}}|F_{j<0}\rangle\otimes|\Psi_{j>0}\rangle+\frac{a_0 b_0}{\sqrt{2}} |\Psi_{j<0}\rangle \otimes |F_{j> 0}\rangle
\end{equation}
We use the orthonormal $|\Psi_{j<0}\rangle$ and $|F_{j<0}\rangle$ to construct an orthonormal basis of the states with all electrons having $j<0$,  and calculate the partial trace of the density matrix. 
\begin{equation}
	\rho_{j>0}=\text{Tr}_{j< 0}|\Psi_{\text{GS}}\rangle\langle\Psi_{\text{GS}}|=a_0^2\begin{pmatrix}1+\frac{b_0^2}{2}&\frac{b_0}{\sqrt{2}}\\\frac{b_0}{\sqrt{2}}&\frac{b_0^2}{2}\end{pmatrix},
\end{equation}
which is Eq. (8) in the main text. The matrix elements of $\rho_{j>0}$ associated with states orthogonal to  $|\Psi_{j>0}\rangle$ and $|F_{j>0}\rangle$ are zero and do not contribute to the entanglement entropy.
\\
\\
The eigenvalues of $\rho_{j>0}$ are:
\begin{equation}
	\lambda_{\pm}=\frac{1+Z \pm \sqrt{2Z+Z^2}}{2(1+Z)}.
\end{equation}
Here we have used $b_0^2=\frac{1}{Z}$ and $a_0^2=\frac{Z}{Z+1}$, where $Z$ is the quasiparticle mass enhancement.

Using Eq. (1) in the main text we obtain the entanglement entropy which depends on the model parameters only through $Z$ as:
\begin{equation}
S_{\uparrow} = - \frac{1+Z-\sqrt{2Z+Z^2}}{2(1+Z)}\log_{2}\left(\frac{1+Z - \sqrt{2Z+Z^2}}{2(1+Z)}\right)-\frac{1+Z + \sqrt{2Z+Z^2}}{2(1+Z)}\log_{2}\left(\frac{1+Z + \sqrt{2Z+Z^2}}{2(1+Z)}\right).
\end{equation}
In the strongly correlated regime $Z\sim 0$, a series expansion leads to $S_{\uparrow}\sim 1-\frac{Z}{\ln(2)}$.

\pagebreak

\section{Accuracy of the Numerical calculations}
In the density matrix renormalization group (DMRG) calculations, there are two relevant parameters to consider: the final size $L$ of the system after the renormalization procedure and the maximum number $m$ of states kept at each iteration step.

\begin{figure}[H]\center
\includegraphics[width=0.5\columnwidth,angle=0,keepaspectratio=true]{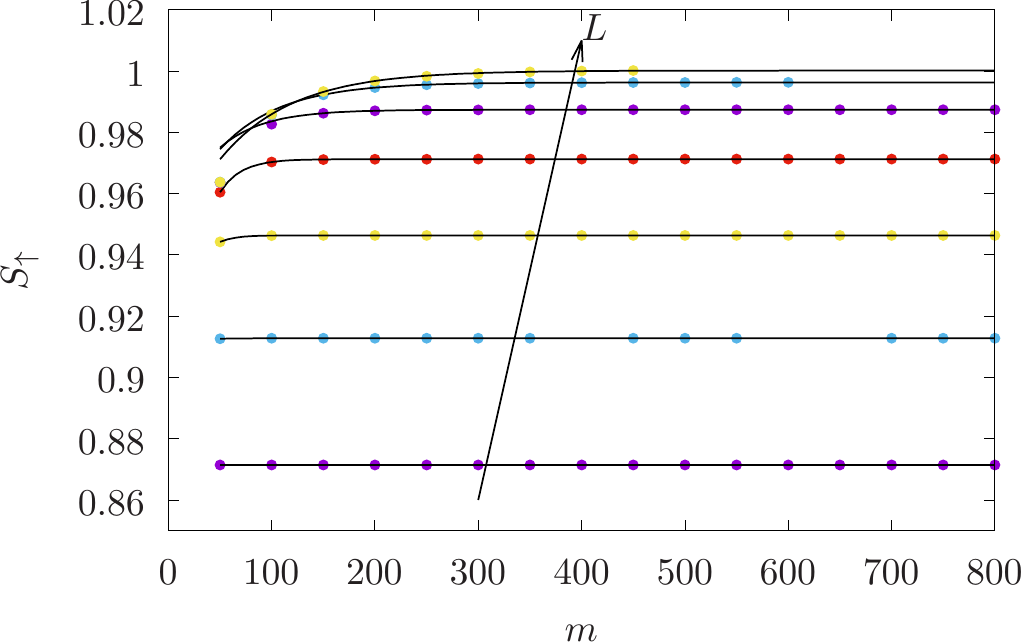}
	\caption{Spin entanglement entropy $S_\uparrow$ in the $N=2$ Anderson model as a function of the number of states $m$ kept at each renormalization group step. The calculation parameters are $U=3D$, $\Gamma=0.32D$, where $D$ is the width of the conduction band. 
	}
	\label{fig:entopml}
\end{figure}

\begin{figure}[H]\center
\includegraphics[width=0.5\columnwidth,angle=0,keepaspectratio=true]{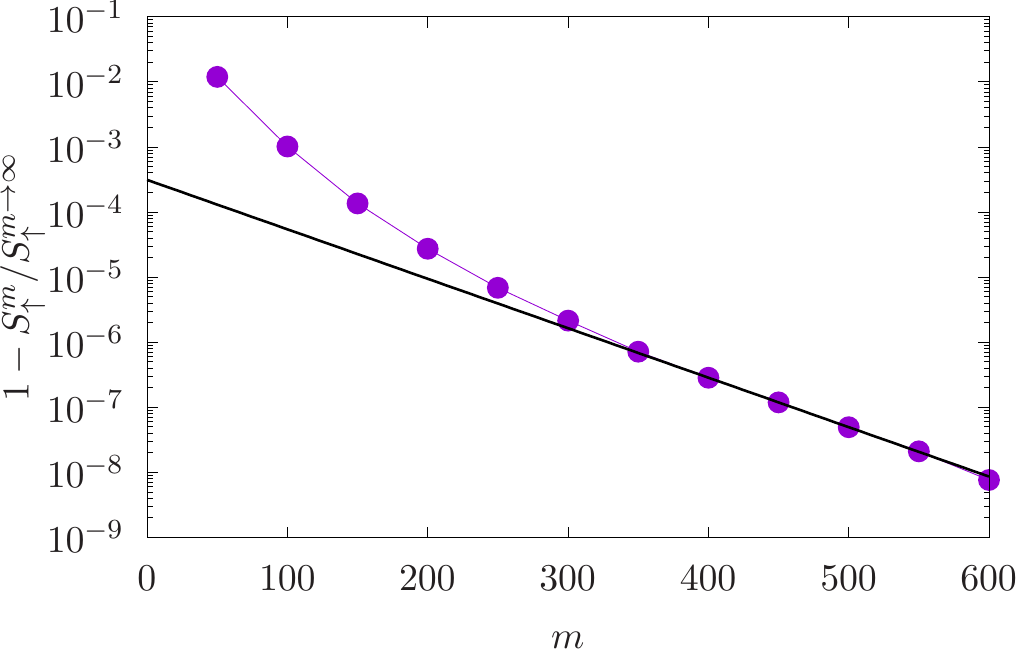}
\caption{Difference between the entropy for a given $m$ with the $m\to \infty$ extrapolated value. The parameters of the Anderson model are $N=2$, $U=4D$, $\Gamma=0.405D$.
	}
	\label{fig:errm}
\end{figure}

\begin{figure}[H]\center
\includegraphics[width=0.5\columnwidth,angle=0,keepaspectratio=true]{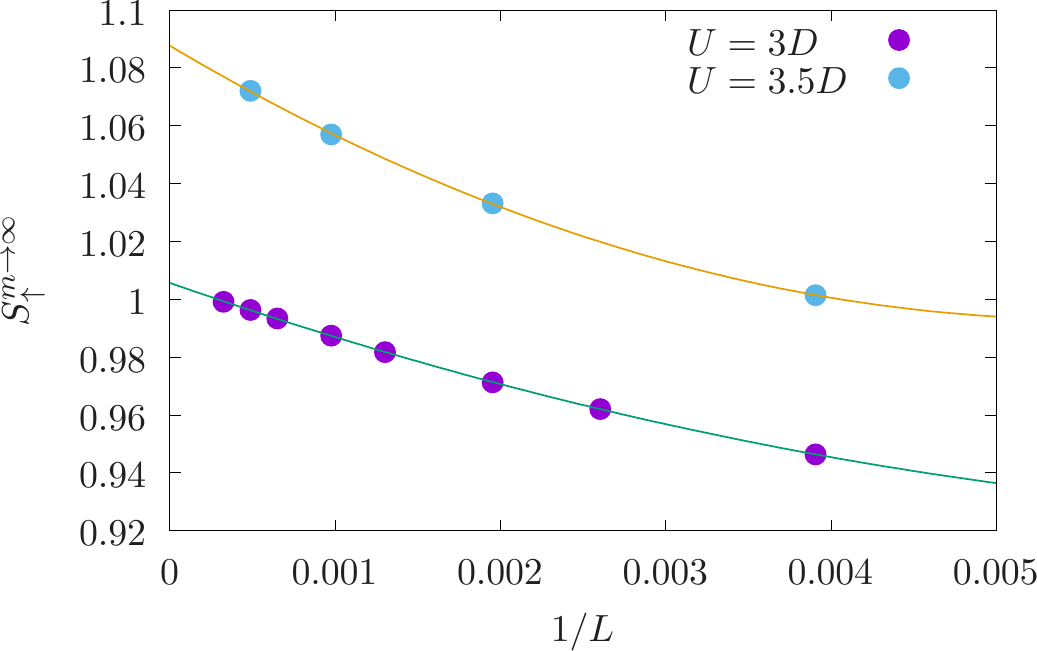}
	\caption{Spin entanglement entropy $S_\uparrow$ as a function of the inverse DMRG chain length $1/L$. The fitting function is of the form $S(1/L)=S(0)+a/L+b/L^2$. The parameters of the Anderson model are $N=2$ and $\Gamma=0.32D$.
	}
	\label{fig:entopm}
\end{figure}

In Fig. \ref{fig:entopml} we present the calculated spin entanglement entropy $S_\uparrow$ in the Anderson model as a function of $m$ and for different values of $L$.  The dots are the numerical data and the lines are least squares fits (see below) using a function of the form $S_\uparrow(m)=S_\uparrow^{m\to\infty}-c_1 e^{-m/c_2}$, with $S_\uparrow^{m\to\infty}$, $c_1$, and $c_2$ the fitting parameters. As it can be seen in Fig. \ref{fig:errm}, $S_\uparrow$ converges exponentially to $S_\uparrow^{m\to\infty}$ with increasing $m$. The $m$-extrapolated values $S_\uparrow^{m\to\infty}$ for each system size $L$ are then extrapolated to the $L\to \infty$ limit, assuming finite size corrections polynomial in $1/L$ (see Fig. \ref{fig:entopm}). For all the parameters presented in the main text the extrapolation corrections amounted to less than 2\%.

\begin{figure}[H]\center
\includegraphics[width=0.5\columnwidth,angle=0,keepaspectratio=true]{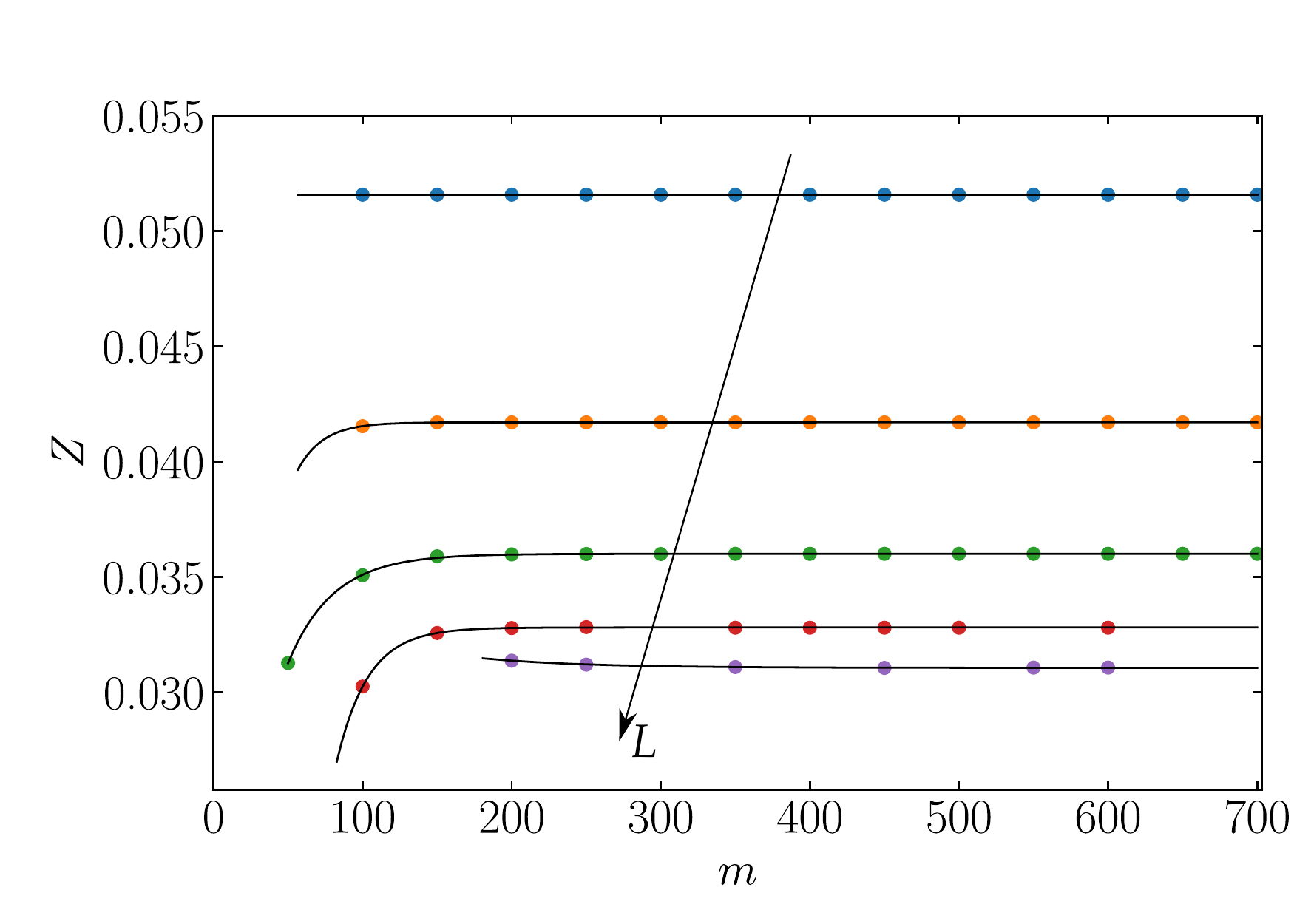}
	\caption{Quasiparticle mass enhancement $Z$ as a function of the maximum number of states $m$ kept at each DMRG step. The fits are a function of an exponential curve. The parameters of the Anderson model are $N=2$ $U=3D$ and $\Gamma=0.32D$.
	}
	\label{fig:sucepm}
\end{figure}

\begin{figure}[H]\center
\includegraphics[width=0.5\columnwidth,angle=0,keepaspectratio=true]{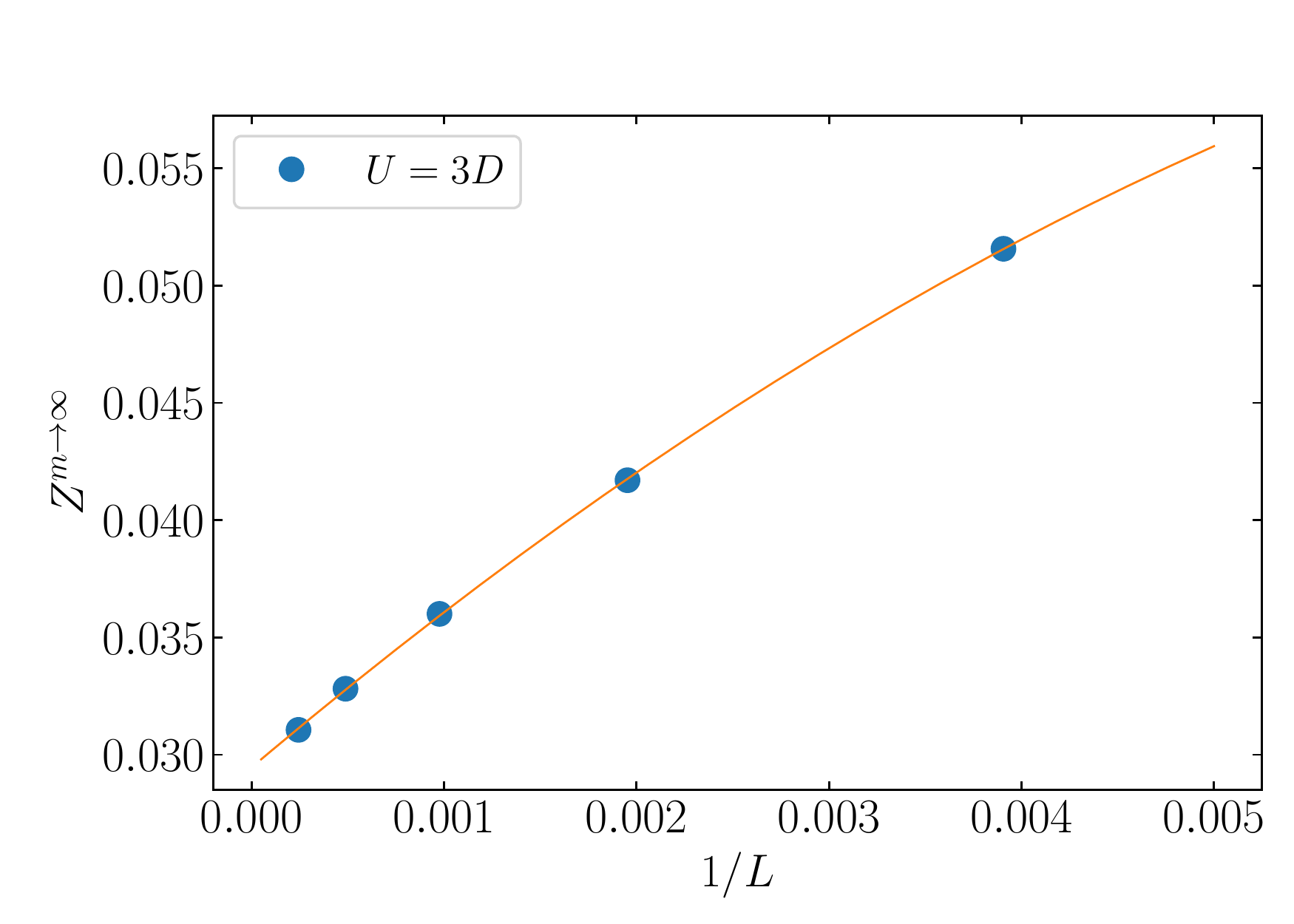}
	\caption{Quasiparticle mass enhancement $Z$ as a function of the inverse DMRG chain size $1/L$. The fitting function is: $Z=Z(0)+a/L+b/L^2$. The parameters of the Anderson model are $N=2$, $\Gamma=0.32D$, and $U=3D$.}
	\label{fig:sucel}
\end{figure}

Finally, we present how the extrapolation of the quasiparticle mass enhancement $Z$ to the $m \to \infty$ and $L\to \infty$ limits was performed (see Fig. \ref{fig:sucepm} and Fig. \ref{fig:sucel}). The quasiparticle mass enhancement is calculated using $Z\sim(\Gamma \chi)^{-1}$, where $\chi$ is the magnetic susceptibility and $\Gamma$ the hybridization between the impurity and the conduction band. The magnetic susceptibility is calculated applying a Zeeman energy splitting $\delta=0.0001D$ at the impurity ($\chi$ is independent of the value of $\delta$ if $\delta$ is smaller than $k_BT_K$ and large enough to avoid numerical precision errors). Fig. \ref{fig:sucepm} presents $Z$ vs. $m$ for different values of $L$, the lines are least squares fits of $Z(m)=Z^{m\to\infty}-b_1 e^{-m/b_2}$, with $Z^{m\to\infty}$, $b_1$, and $b_2$ the fitting parameters. In Fig. \ref{fig:sucel}, the $m$-extrapolated values of $Z$ for each DMRG chain size $L$ are fitted using a second degree polynomial in $1/L$.

\pagebreak
\pagebreak

\section{Large-$\mathcal{J}$ perturbation theory for the ground state wave function}
In the large-$\mathcal{J}$ limit the impurity and the first site of the tight-binding chain form a strong spin singlet and effectively decouple from the rest of the tight-binding chain.
The ground state of the system is a direct product of this spin singlet $|S\rangle$ and a filled Fermi sea for the rest of the chain $|\Omega\rangle$. The spin entanglement in this case is given by the spin singlet and is simply $S_\uparrow(\mathcal{J}\to\infty)=1$. To obtain the ground state in perturbation theory for finite $t/\mathcal{J}$ it is convenient to write the Hamiltonian as $H=H_K+H^\prime+H_{e}$ where
\begin{equation}
H_K=\mathcal{J} {\bm S}_f\cdot {\bm S}_1,
\end{equation}	
\begin{equation}
	H_{e} = - t \sum_{i=2}^L\sum_{j=\{\uparrow,\downarrow\}} \left(c_{i,j}^\dagger c_{i+1,j}^{} + \text{H.c.}\right),
	\label{eq:tbchain}
\end{equation}	
and consider 
\begin{equation}
	H^\prime = - t \sum_{j=\{\uparrow,\downarrow\}} \left(c_{1,j}^\dagger c_{2,j}^{} + \text{H.c.}\right)
	\label{eq:tbchain}
\end{equation}	
as a perturbation.
To first order in $t/\mathcal{J}$ the ground state wavefunction is:
\begin{equation}\begin{split}
|\Psi_\text{GS}\rangle^{(1)} &=|\Psi_\text{GS}\rangle^{(0)}-\frac{t}{\sqrt{L}}\sum_{k} \frac{\sin(k a)}{-\frac{3}{4}\mathcal{J}-E_k}\left(c_{f\uparrow}^\dagger c_{k\downarrow}^\dagger-c_{f\downarrow}^\dagger c_{k\uparrow}^\dagger\right)|\Omega\rangle\\
&+\frac{t}{\sqrt{L}}\sum_{k} \frac{\sin(k a)}{-\frac{3}{4}\mathcal{J}+E_k}\left(c_{f\uparrow}^\dagger c_{1\uparrow}^\dagger c_{1\downarrow}^\dagger c_{k\uparrow}+c_{f\downarrow}^\dagger c_{1\uparrow}^\dagger c_{1\downarrow}^\dagger c_{k\downarrow}\right)|\Omega\rangle,       
\end{split}
\end{equation}
with $k=\frac{n\pi}{aL}$  and $n =\{ 1,2, ... ,L-2,L-1\}$. To calculate the spin entanglement entropy it is convenient to define the operators:
\begin{equation}
	\tilde{c}_{\alpha,j}=\frac{1}{\lambda_\alpha}\frac{2}{\sqrt{L}} \sum_{k }^{\text{ occ}} \frac{\sin(k a)}{1-\frac{4 E_k}{3\mathcal{J}}}c_{k,j}
\end{equation}
and
\begin{equation}
	\tilde{c}_{\beta,j}^\dagger=\frac{1}{\lambda_\beta}\frac{2}{\sqrt{L}} \sum_{k }^{\text{ unocc}} \frac{\sin(k a)}{1+\frac{4 E_k}{3\mathcal{J}}}c_{k,j}^\dagger
\end{equation}
where the sum over $k$ is restricted to states occupied (unoccupied) in $|\Omega\rangle$ and $\lambda\sim 1$ is a normalizing constant.  For the half-filled case (electron-hole symmetry) we have $\lambda_\alpha=\lambda_\beta=\lambda$ and $\lambda=1$ to lowest order in $t/\mathcal{J}$.
Using these operators to construct a basis, the reduced density matrix reads
\begin{equation}
	\rho_\uparrow=\frac{1}{1+\left(\frac{4 t \lambda}{3\mathcal{J}}\right)^2}\begin{pmatrix}\frac{1}{2}+\frac{4}{9}\frac{t^2}{\mathcal{J}^2}\lambda^2&\frac{\sqrt{2}}{3}\frac{t}{\mathcal{J}}\lambda&0&0\\\frac{\sqrt{2}}{3}\frac{t}{\mathcal{J}}\lambda &\frac{4}{9}\frac{t^2}{\mathcal{J}^2}&0&0\\
	0&0&\frac{1}{2}+\frac{4}{9}\frac{t^2}{\mathcal{J}^2}\lambda^2&\frac{\sqrt{2}}{3}\frac{t}{\mathcal{J}}\lambda\\0&0&\frac{\sqrt{2}}{3}\frac{t}{\mathcal{J}}\lambda &\frac{4}{9}\frac{t^2}{\mathcal{J}^2}\end{pmatrix}
\end{equation}
which has double degenerate eigenvalues: $\left\{2 \left(\frac{2}{3}\frac{t}{\mathcal{J}}\right)^4,\,\frac{1}{2}-2 \left(\frac{2}{3}\frac{t}{\mathcal{J}}\right)^4\right\}$, to lowest non-trivial order in $t/\mathcal{J}$. 
The entanglement entropy to lowest non-trivial order in $D/\mathcal{J}$ ($D=2t$ is the half-bandwidth) reads:
\begin{equation}
  S_\uparrow\sim 1 + \frac{2}{81} \left(\frac{D}{\mathcal{J}}\right)^4 \log_2(\mathcal{J}/D)
\end{equation}

Considering that we started from a first order correction to the wavefunction, 
higher order corrections in the wave function would be needed to obtain the exact expression for the eigenvalues to fourth order. However, the expression $S_\uparrow=1 + c_0 \left(\frac{D}{\mathcal{J}}\right)^4 \log_2(\mathcal{J}/D)$ provides an excellent fit (using $c_0$ as a fitting parameter) to the numerical data (see main text).

\bibliographystyle{apsrev4-1}
\bibliography{references}